\documentclass[prb,floats,epsf,twocolumn]{revtex4}
 \usepackage{graphics}
\begin{document}

\title{Relativistic Mott criticality in graphene}

\author{Igor F. Herbut$^1$, Vladimir  Juri\v ci\' c$^1$, and Oskar Vafek$^2$}

\affiliation{$^1$ Department of Physics, Simon Fraser University,
 Burnaby, British Columbia, Canada V5A 1S6 \\
 $^2$ National High Magnetic Field Laboratory and Department of Physics, Florida State University, Tallahassee, Florida 32306, USA}

\begin{abstract}
We formulate the effective Gross-Neveu-Yukawa theory of the
semimetal-insulator transitions on the honeycomb lattice and compute
its quantum critical behavior near three (spatial) dimensions. We
find that at the critical point Dirac fermions do not survive as
coherent excitations and that the $\sim 1/r$ tail of the weak Coulomb
interaction is an irrelevant coupling. The emergent Lorentz
invariance near criticality implies a universal ratio of the
low-temperature specific heats of the metallic and the
rotational-symmetry-broken insulating phase.
\end{abstract}
\maketitle

\vspace{10pt}

\section{Introduction}

It is well known that the large overlap between the $p_z$ orbitals from neighboring carbon atoms makes graphene
an excellent conductor. \cite{rmp} Nevertheless, one can conceive of situations in which the relative strength
of the repulsive Coulomb interaction between electrons would be higher so that graphene would turn into a Mott
insulator. A recent calculation suggests that just taking the graphene sheet away from the substrate may gap
out the Dirac points. \cite{drut,vafekcase} If so, varying the dielectric constant of the surrounding medium
over a sufficiently wide range could in principle be used to tune through the metal-insulator (MI) transition
in graphene. Another possibility would be stretching the sheet to reduce the hopping between the $p_z$
orbitals. Such a quantum phase transition would be the analog of the Higgs mechanism for gauge-neutral fermions
in particle physics, with the Higgs boson here as a composite field. It would represent maybe the simplest
example of fermionic quantum criticality, in which the gapless fermions exist only near the isolated points in
the momentum space. A finite gap would also make graphene more interesting for potential applications in
electronics. \cite{geim}

 A unique feature of the Mott phase in graphene is that it may come in several varieties:
 from the familiar N\'{e}el and
 staggered-density phases,
 \cite{semenoff, herbut1} to the more exotic insulators that break the time-reversal symmetry (TRS).
 \cite{herb-jur-roy} There have been several studies of the MI transition in graphene.
 \cite{drut, herbut1, herb-jur-roy, sorella, martelo, paiva, gorbar, leal}
 Nevertheless, several fundamental questions still await answers.
 Among these, the following qualitative issues seem particularly pertinent:
 1) what is the role of Dirac fermions in the critical behavior,
 2) how does the criticality depend on the nature of the order parameter (OP) in the Mott insulator,
 3) what is the fate of fermions near the critical point, 4)
 is the long-range tail of the Coulomb interaction relevant?
 Questions 1) and 3) echo some of the central themes of the wider field of quantum critical phenomena,
 \cite{vojta, sachdev} whereas the last question, as we will argue, is related to the classic
 problem of triviality of the continuum limit in non-asymptotically free field theories.

   In this paper we present an effective theory of the Mott transition in graphene which allows
   us to address these and related issues in a controlled and transparent way.
   Our action contains both the self-interacting bosonic (or "Higgs") OPs
   and the Dirac fermions, coupled by Yukawa-like terms.
   It represents a simple modification of the Gross-Neveu theory, derived previously in the large-N limit,
   \cite{herbut1} but with a crucial novel feature:
   there is an upper critical (space) dimension in the problem of {\it three}.
   This allows one to perform the $\epsilon=3-d$ expansion, with the Higgs and the fermionic fields
   at all stages of the calculation treated on the same footing, thus placing the Mott criticality in graphene
   at the same level of rigor as the textbook $\Phi^4$ theory. We find the MI transition in graphene
   to be of the {\it second order}, and to be governed by the critical point laying at a finite Yukawa coupling (Fig. 1). Although crucial for the critical behavior, Dirac fermions acquire a small positive anomalous dimension, so that the residue of the quasiparticle pole continuously vanishes as the transition is approached from the metallic side. Whereas the transition may be tuned by increasing the strength of Coulomb repulsion, its $\sim 1/r$ tail is in fact an {\it irrelevant} perturbation to the leading order in $\epsilon$. We determine the dependence of the critical exponents on the broken symmetry of the Mott phase.
   Our analytical results compare favorably with those of a recent numerical work. \cite{drut}
   The emergent Lorentz symmetry  near criticality implies the existence of a new universal quantity: the ratio of the low-temperature specific heats of the semimetal and of the rotationally non-invariant insulator. Finally, possible
   analogies between the Mott criticality in graphene and the chiral symmetry breaking in $3+1$-dimensional quantum electrodynamics ($QED_4$) are noted.

   The remainder of the paper is organized as follows.
   In the next section we discuss the order parameters and the effective field theory near the transition.
   In sec. III we perform the analysis of the problem near 3+1 dimensions,
    and in sec. IV we discuss our results, as well as the universal amplitudes specific to the problem at hand.

\section{Effective theory}

Let us begin by fixing the necessary notation. Introduce the Dirac fermion for spin-1/2 electrons as $\Psi^\dagger = (\Psi_\uparrow ^\dagger, \Psi_\downarrow ^\dagger)$,  where each spin component is defined as
\begin{widetext}
\begin{equation}\label{Dirac-spinor}
\Psi_\sigma ^\dagger(\vec{x},\tau)=T{\sum_{\omega_n}}{\int^\Lambda}{\frac{d\vec{q}}{(2\pi
a)^2}}e^{{i{\omega_n}\tau}+i{\vec{q}\cdot{\vec{x}}}}(u_\sigma^\dagger(\vec{K}+\vec{q},\omega_n),
v_\sigma^\dagger(\vec{K}+\vec{q},\omega_n),
u_\sigma^\dagger(-\vec{K}+\vec{q},\omega_n),
v_\sigma
^\dagger(-\vec{K}+\vec{q},\omega_n)),
\end{equation}
\end{widetext}
where $\vec{K}= (1,1/\sqrt{3}) (2 \pi/a\sqrt{3})$ is the Dirac point,
and $u$ and $v$ are the Grassmann fields on the two sublattices of the honeycomb lattice.
The reference frame is rotated so that $q_x=\vec{q}\cdot\vec{K}/K$.
The non-interacting Lagrangian is then  $L_f = \bar{\Psi}\gamma_\mu \partial_\mu \Psi$,
with the $\gamma$-matrices defined as $\gamma_0= I_2 \otimes \sigma_z$, $\gamma_1= \sigma_z \otimes \sigma_y$,
and $\gamma_2= I_2 \otimes \sigma_x$, and $\bar{\Psi}=\Psi^\dagger \gamma_0$.
The ultraviolet cutoff is $\Lambda\ll 1/a$, where $a$ is the lattice spacing.
We have set the Fermi velocity to unity.

\begin{figure}[t]
{\centering\resizebox*{75mm}{!}{\includegraphics{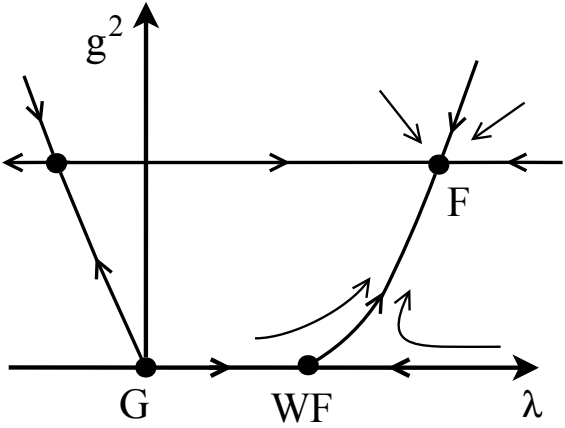}}
\par} \caption[] {Schematic flow in the critical plane for Mott transitions in graphene.
For any positive bosonic quartic coupling $\lambda$ the transition is continuous
and governed by the fermionic critical point F. $g$ is the Yukawa coupling between the OP and Dirac fermions.
The other marked fixed points are the Gaussian (G) and the Wilson-Fisher (WF). The (unmarked) bicritical
fixed point is unphysical, as argued below. }
\end{figure}

In 2+1 dimensions there are two qualitatively different ways to gap out the Dirac fermions. \cite{herb-jur-roy}
The first one preserves the chiral symmetry (CS) generated by $\{ \gamma_3, \gamma_5, \gamma_{35} \}$,
$\gamma_{35} = i\gamma_3 \gamma_5$, but breaks the TRS for each spin component separately. The time-reversal in
the above representation is defined as  $\Psi_\sigma \rightarrow I_t \Psi_\sigma$, where $I_t= (\sigma_x
\otimes I_2) K$ and  $K$ is the complex conjugation. \cite{herbut2}
Choosing  $\gamma_3= \sigma_x \otimes
\sigma_y$ and $\gamma_5= \sigma_y \otimes \sigma_y$, the TRS-breaking OP may be written as

   \begin{equation}
   \phi= (\phi_s, \vec{\phi}_t) = (\langle \bar{\Psi} \gamma_{35} \Psi \rangle, \langle \bar{\Psi} \vec{\sigma}
   \gamma_{35} \Psi \rangle),
   \end{equation}
   with $\phi_s$ and $\phi_t$ as its (real) singlet and triplet components.
   The Pauli matrices act on spin, and $\gamma$-matrices act on Dirac indices.
   A lattice realization of $\phi$ has been offered long ago in terms of circulating currents
   between the neighboring sites on the same sublattice. \cite{haldane}
   It was also shown recently that $\phi\neq 0$ is favored by the second nearest-neighbor
   repulsion on honeycomb lattice. \cite{raghu}

  The second order parameter preserves the TRS, but breaks the CS:
  \begin{equation}
  \chi=  (\chi_s, \vec{\chi}_t) = (\langle \bar{\Psi} \Psi \rangle,  \langle \bar{\Psi} \vec{\sigma}\Psi \rangle ).
   \end{equation}
 In the representation above, for example, the singlet corresponds to staggered density,
 and the triplet to staggered magnetization. A finite $\chi_s $  ($\chi_t$)
 may then be induced by a large nearest-neighbor (on-site) repulsion. \cite{herbut1}
 In this case a finite singlet component breaks the Ising symmetry of the sublattice-exchange,
 whereas the triplet breaks also the (spin) rotational symmetry.

  We may promote the above fermion bilinears into separate dynamical fields and write the effective
  action near the Mott transition as $S=\int d\tau d\vec{x} L$, with $L=L_f + L_y + L_b+ L_c$,
  where the bosonic part is
 \begin{widetext}
  \begin{equation}
 L_b =  \sum_{\psi = \phi,\chi } \{ \frac{1}{2} [  \psi_s (-\partial_\tau ^2 -
 v _{\psi,s} ^2 \nabla^2  + t_{\psi,s})\psi_s
 +  \vec{\psi}_t \cdot (-\partial_\tau ^2 - v _{\psi,t} ^2 \nabla^2   + t_{\psi,t}) \vec{\psi}_t ] +
 \lambda_{\psi,s} \psi_s ^4 + \lambda_{\psi,t} (\vec{\psi}_t \cdot \vec{\psi}_t )^2  + \lambda_{\psi, st}
 \psi_s ^2 \vec{\psi}_t ^2 \},
 \end{equation}
 \end{widetext}
 and the Yukawa terms that couple bosonic and fermionic fields are
 \begin{equation}
 L_y = \sum_{\psi=\chi,\phi} \{ g_{\psi,s} \psi_s \bar{\Psi} M_\psi \Psi +
  g_{\psi,t} \vec{\psi}_t \cdot \bar{\Psi} M_\psi \vec{\sigma} \Psi \},
 \end{equation}
 where the matrices $M_\chi = 1$, and $M_\phi=\gamma_{35}$. Note that $L$ lacks
 the Lorentz symmetry due to the assumed generic bosonic velocities $v_{\psi,s}\neq v_{\psi,t}\neq 1$.
 The form of $L$  is dictated by the spin-rotational, time-reversal, and sublattice exchange (Ising)
 symmetry of the original problem on honeycomb lattice.

 One may also account for the long-range tail of the Coulomb interaction by including the term
 \begin{equation}
 L_c = i a \bar{\Psi} \gamma_0 \Psi + \frac{1}{2 e^2} a |\nabla|^{d-1} a.
 \end{equation}
 We allow for a general spatial dimension $d$, but in a manner that assures that the
 integration over the scalar gauge field $a$ would introduce the $\sim e^2 /r$
 density-density interaction between fermions in any $d$. \cite{herbut3}

    Integrating out the Higgs OPs in the symmetric phase $t_{\psi,s}, t_{\psi,t} >0$
    would produce local quartic terms for the remaining fermionic degrees of freedom,
    and reduce the theory to the Gross-Neveu form derived before in the large-N limit. \cite{herbut1}
    The advantage of the Yukawa form is that it becomes renormalizable in $3+1$ dimensions, where {\it both}
    the Yukawa and the Higgs self-interaction couplings become dimensionless. The relationship between the
    Gross-Neveu and the Yukawa descriptions mirrors the one between the non-linear and linear sigma-models
    in standard critical phenomena, \cite{zinn} and there is a substantial evidence that both describe the
    same critical point. \cite{wetterich}
    In the remaining part of the paper we exploit this equivalence
    and study the effective Yukawa theory in $d=3-\epsilon$ dimensions, with an eye on  $\epsilon=1$
    relevant to graphene.

    A possible objection to the  pursuit of the above strategy may be that the Higgs bosons $\phi$
    that involve $\gamma_{35}$ do not gap the Dirac fermions in $3+1$ dimensions, where the CS is
    reduced to $U(1)$. If we were to compute the beta-functions directly in $2+1$ dimensions, on the other hand,
    at least perturbatively there can be no difference between those with $\chi$ and $\phi$ labels.
    This is because the matrix $\gamma_{35}$ commutes with all the $\gamma$-matrices which appear in
    the fermion propagators. \cite{herb-jur-roy} This leads us to
    conjecture that the critical behaviors at $\chi$ and $\phi$ transitions that share the same
    spin-symmetry are identical.

\section{$\epsilon$ - expansion}
    With the above rational in mind, hereafter we consider only the CS breaking transition.
      Anticipating some of the results we set $v_{\chi,s}= v_{\chi,t}=1$ and $e^2 =0$,
      and integrate out both the fermionic and the bosonic modes within the four-momentum
      shell $\Lambda/b < (\omega^2 + \vec{k}^2)^{1/2}  < \Lambda$. There are two critical
      points, distinguished by the rotational symmetry of the OP: 1) at $t_{\chi,s}=0$
      ($t_{\chi,t}=\infty$), and 2) at $t_{\chi,t}=0$ ($t_{\chi,s}=\infty$). \cite{remark}
      They may be tuned by increasing the nearest-neighbor and the on-site
      repulsive terms in the Hubbard model, respectively. \cite{herbut1} To one-loop order we find
      \begin{equation}
      \frac{dg^2 _{u} }{d\ln b}= g^2 _{u} (\epsilon - (7-S)  g^2 _{u}),
      \end{equation}
      \begin{equation}
      \frac{ d\lambda_{u} }{ d\ln b}= \lambda_{u}(\epsilon - 8 g^2 _{u}) - 4 (9 + S) \lambda^2 _{u}+ 2 g^4 _{u},
    \end{equation}
    where $g_u = g_{\chi,u}/(8\pi^2\Lambda^{\epsilon})$ and $\lambda_u = \lambda_{\chi,u} /(8\pi^2 \Lambda^{\epsilon})$,
    $u=s,t$, and $S=0$ ($S=2$) for singlet (triplet) couplings. The purely bosonic Wilson-Fisher fixed point
    is unstable in the $g$-direction, and the critical point lies at a finite Yukawa and quartic couplings (Fig. 1).
     The flows in both cases resemble the one of superconductors, \cite{herb-tes} except that the
     bicritical point now lies in the unphysical region, so that both the singlet
     and the triplet transition are always second-order.

   Let us examine the stability of the critical point with respect to perturbations.
   First, breaking the Lorentz symmetry by a small difference in the fermionic and
   bosonic velocities is irrelevant: for $\delta_{u}=1-v_{\chi,u} \ll 1$, to the leading order in $\epsilon$ we find
      \begin{equation}
      \frac{d\delta _{u}} {d\ln b}= \frac{4\epsilon}{S-7} \delta_{u}.
      \end{equation}
      Weak Coulomb interaction is also readily found to be (marginally) irrelevant
      \begin{equation}
      \frac{de^2}{d\ln b} = - \frac{4}{3}( 2 \delta_{d,3} + 1) e^4,
      \end{equation}
      similarly as in the purely bosonic theories.  \cite{herbut3} Here we rescaled $e^2 /8 \pi^2 \rightarrow e^2$.
      The first term in the bracket derives from the usual polarization, and it is present only in $d=3$,
      where the second term in Eq. (6) becomes analytic in momentum. \cite{herbut3}
      The remaining term comes from the renormalization of the  Fermi velocity. \cite{gonzales}

       The correlation length critical exponent is then
      \begin{equation}
      \nu= \frac{1}{2}+ \frac{3 (5+S)}{(7-S)(9+S)} \epsilon + O(\epsilon^2),
      \end{equation}
      and
      \begin{equation}
      \eta_b= \frac{4}{7-S} \epsilon +  O(\epsilon^2),
      \end{equation}
      with $\eta_b$ as the usual OP's anomalous dimension.

The fermion propagator at the critical point behaves as $G_f ^{-1} \sim (\omega^2 + k^2)^{  (1-\eta_f) / 2}$,
where the {\it fermionic} anomalous dimension is
 \begin{equation}
 \eta_{f}= \frac{ 3}{ 2(7-S)}  \epsilon + O(\epsilon^2).
 \end{equation}
 The scaling then implies \cite{herbut1} that the residue of the Dirac quasiparticle's pole
 vanishes near the critical point as a power-law,
 \begin{equation}
 Z\sim (t_{\chi,u}) ^{ 3 \epsilon /(4(7-S)) + O(\epsilon^2)}.
 \end{equation}
There are therefore no sharp fermionic excitations right at the critical point.

   Both the bosonic and the fermionic anomalous dimensions, being proportional to the critical point
   values of the Yukawa coupling, are finite already to the leading order in $\epsilon$.
   This implies that the exponents have rather different values from the usual mean-field.
   Since there does not seem to be a dangerously irrelevant coupling in the problem the
   hyperscaling should hold and we may obtain the remaining exponents from the usual scaling relations.
   \cite{book} This way we find
   \begin{equation}
   \gamma= 1 + \frac{4(3+S)}{(7-S)(9+S)} \epsilon +O(\epsilon^2),
   \end{equation}
   \begin{equation}
   \delta= \frac{d+3-\eta_b}{d-1+\eta_b} = 3-\frac{1+S}{7-S} \epsilon +O(\epsilon^2).
   \end{equation}
   For a large number of Dirac fields $N$ one similarly finds: \cite{herbut1} $\nu=1+O(1/N)$
   and $\eta_b= 1+ O(1/N)$, and thus $\gamma= 1+ O(1/N)$, and $\delta= 2 + O(1/N)$.
   Various calculations on the Gross-Neveu models in the past \cite{wetterich} also
    suggested that the infinite-N results are often good estimates of the exponents' actual values.

\section{Discussion}

     The critical exponents have recently been computed numerically for two species of Dirac fermions
     interacting via $\sim 1/r$ interactions. Whereas only the standard power-laws and no essential
     singularity \cite{gorbar,leal} were observed, \cite{drut} the critical exponents $\gamma=1$,
     and $\delta\approx 2.3$, particularly when expressed in terms of $\nu=0.85$ and $\eta_b= 0.82$
     appear distinctly not to be of the usual mean-field variety. They do seem close to our one-loop results,
     however, particularly for the triplet OP, where for $\epsilon=1$ we find $\nu\approx 0.88$
     and $\eta_b\approx 0.80$.
     The correlation-length exponent is also very close to the previous two-loop result \cite{rosenstein}
     for the singlet OP.  These are only crude estimates, but we believe that the observation of the values of
     $\nu$ and $\eta_b$ near unity should be taken as a sign of the Gross-Neveu-Yukawa fermionic criticality.
     Although the strength of $1/r$ interaction may be used as the tuning parameter, both the present $\epsilon$-,
     earlier \cite{herbut1, aleiner, son} $1/N$-expansions, and the results near $1+1$ dimensions \cite{semenoff1}
     suggest this to be ultimately an irrelevant coupling. The Mott transition in graphene
     may be analogous to the situation in the  $QED_4$, \cite{kogut}
     where the CS breaking transition brought by the increase of the electromagnetic
     charge appears to be described by the Nambu-Jona-Lasinio Lagrangian with only short-range interactions.
     This is one way to phrase the triviality of the continuum limit of $QED_4$. \cite{gockeler}

Let us list some further results specific to the Mott criticality in graphene.
 As the critical point is approached from the insulating side both the masses of the Higgs OP
 and of the Dirac fermions approach zero, with the universal ratio
       \begin{equation}
       (\frac{m_b}{m_f})^2 = \frac{ 8\lambda_u }{g_u ^2} = \frac{16}{9+S} + O(\epsilon),
       \end{equation}
analogous to the universal Ginzburg-Landau parameter at the superconducting critical point. \cite{herb-tes,
book} The emerging Lorentz invariance in the critical region implies that a) the dynamical critical exponent
$z=1$, so that the Fermi velocity is not critical, \cite{herb-jur-roy} b) the (non-universal) velocity of
bosonic excitations in the critical region approaches the (non-universal) Fermi velocity. For the triplet OP
this means that the spin-waves on the insulating side will have the same velocity as the Dirac fermions on the
 metallic side, when both are near criticality. As a consequence, the ratio of the specific heats on the
 metallic and insulating side in the critical region approaches a {\it universal} value
\begin{equation}
\lim_{T \rightarrow 0} \frac{C(t_{\chi,t} \rightarrow 0+) }{C(t_{\chi,t}\rightarrow 0-) }
= 4(1-2^{-d}).
\end{equation}
The numerator (denominator) corresponds to the specific heat of the eight (two) {\it free}
relativistic fermions (Goldstone  bosons) in $d$ dimensions. Assuming that the Lorentz invariance
 is still emergent at the critical point in $d=2$  \cite{herb-jur-roy} the same ratio is exactly
 $3$ at the transition into the spin-density-wave phase in the Hubbard model on honeycomb lattice.

 In conclusion, we presented an effective theory of the Mott transition on honeycomb lattice
 that becomes solvable near three spatial dimensions. We argued that the quantum critical point
 in graphene should be relativistically invariant,  and possibly logarithmically trivial.
 An interesting consequence of our theory is the existence of a simple universal ratio of
 the specific heats of the two phases near  the transition. This particular prediction should
 hopefully become testable in the future numerical and experimental studies.

\section{Acknowledgement}

  I. F. H. and V. J. are supported by the NSERC of Canada.


\begin{thebibliography}{99}
\bibitem{remark} Here we did not consider multicritical behavior and thus set $\lambda_{\psi,st}=0$.
\end{thebibliography}
\end{document}